\documentclass[journal ]{new-aiaa}
\usepackage[utf8]{inputenc}
\usepackage{textcomp}
\usepackage{siunitx}
\usepackage{subcaption}
\usepackage{amsmath}
\usepackage[version=4]{mhchem}
\usepackage{siunitx}
\usepackage{svg}
\usepackage{graphicx}
\usepackage{epstopdf}
\usepackage{longtable,tabularx}
\title{Deep Reinforcement Learning for Separation Control in Turbulent Wind-Tunnel Flow}

\author{Sofia Avdiiv \footnote{Graduate Student.}}
\affil{Technische Universität Berlin, 10587 Berlin, Germany}
\author{Andre Weiner\footnote{Postdoctoral researcher.}}
\affil{Technische Universität Dresden, 01069 Dresden, Germany}
\author{Ben Steinfurth\footnote{Postdoctoral researcher.}}
\affil{Technische Universität Berlin, 10587 Berlin, Germany}

\begin{document}

\maketitle

\begin{abstract}
This work investigates Deep Reinforcement Learning (DRL) as a tool for model-free closed-loop active separation control in a fully turbulent wind tunnel flow over a one-sided diffuser. The agent controls an array of magnetic valves (on/off) that eject compressed air into the boundary layer, while the environmental state is reduced to the signal from a single wall-shear-stress sensor placed near the natural transitory detachment point. The control law is learned in real time using Proximal Policy Optimization. Compared to the standard learning design based on the weighted sum of all rewards following an action, we demonstrate that a horizon aligned with the convective time of the flow leads to faster convergence and a more robust control strategy. The resulting control law corresponds to a low-duty-cycle actuation pattern that yields a forward-flow fraction of approximately 53\%. This compares favorably with conventional and optimized periodic open-loop control ($\sim 40\%$ and $\sim 51\%$, respectively). The findings of this article indicate that, when embedded into an online experiment, DRL represents an efficient tool to identify robust and interpretable active separation control strategies.
\end{abstract}

\section*{Nomenclature}


{\renewcommand\arraystretch{1.0}
\noindent\begin{longtable*}{@{}l @{\quad=\quad} l@{}}

\multicolumn{2}{@{}l}{Experimental Set-Up}\\
$\alpha$ & Diffuser ramp inclination angle, [°]  \\
$\gamma^+$ & Local forward flow fraction, [-] \\
$\theta$ & Boundary layer momentum thickness, [m] \\
$\tau$ & Wall shear stress, [Pa] \\
$\varphi$ & Emission angle, [°] \\
\multicolumn{2}{@{}l}{DRL Set-Up}\\
d$t$ & time step, [s] \\
$\beta$ & Entropy regularization factor, [-] \\
$\gamma$ & Reward discount factor, [-] \\
$\epsilon$ & Policy clip factor, [-] \\
$\pi$ & Policy \\
$\psi$ & Critic weights \\
$\xi$ & Actor weights \\
$\mathcal{A}$ & Advantage\\
$\mathcal{R}$ & Cumulative Reward, Return\\
$\mathcal{V}$ & Value function\\
$\mathcal{L}$ & Loss\\
$\mathcal{H}$ & Entropy\\
$A$ & Action space\\
$S$ & State space\\
$a$ & Action\\
$s$ & State\\
$r$ & Reward\\
$T$ & Trajectory\\
\multicolumn{2}{@{}l}{Operators}\\
$\Delta$ & Difference \\
$\mathbb{E}$ & Expectation \\
$\nabla$ & Gradient \\
\end{longtable*}}
\addtocounter{table}{-1}

\section{Introduction}

Flow separation is often associated with adversities of various kinds, including, but not limited to, loss of aerodynamic lift and increase of drag. Altering the location of the detachment point, reducing the extent of separation, or completely avoiding it are, therefore, the main objectives of flow separation control. These can be achieved through passive or active means, where passive control requires no additional energy source but deliberate (mostly) permanent alterations of geometry or surface treatments, while active control allows for more flexible, targeted manipulation of the flow, but inherently requires additional energy expenditure. 

Flow separation in engineering systems can often be traced back to an adverse pressure gradient, which reduces the momentum flux in the boundary layer. Accordingly, active flow control methods aim to re-energize the boundary layer by adding momentum \cite{lachmann1961boundary, gadelhak1991separation}, for example, through mass flow manipulation in the recirculation region (blowing/suction) by means of synthetic jets \cite{smith1998synthetic, glezer2002synthetic}, sweeping jets \cite{raghu2013fluidic, gregory2013review, ostermann2019interaction}, and pulse jet actuators \cite{petz2007active, warsop2007pulsed, arwatz2008suction, barros2016bluff}, or in the case of plasma actuators \cite{font2006plasma, corke2010plasma, abe2008plasma}, by inducing a body force that generates a near-wall jet without net mass addition. A persistent issue in the design of active flow control (AFC) or active separation control (ASC) strategies is the associated energy consumption, which often outweighs the benefits achieved through their application and compromises their efficiency \cite{intro_gad-el-hak}.

Active control can be further divided into open-loop and closed-loop approaches. A commonly used baseline open-loop strategy is periodic flow excitation with a prescribed forcing signal \cite{loeffler2021fluidic, greenblatt_wygnanski_2000, seifert1996periodic}. Although open-loop strategies can be highly effective, their performance is inherently limited by the absence of feedback \cite{beaudoin2006adaptive}. Generally, designing a feedback control law for ASC is considered a challenge due to the high-dimensional, chaotic nature of the turbulent boundary layer. There are model-based and model-free methods of designing a control law: model-based methods generally rely on physics-informed mathematical models of the system, often applying systemic model-reductions to optimize computational effort \cite{henning2007robust,brandt2011effect, gautier2013pulsed, rowley2017model}; while model-free methods are independent of the mathematical modeling of the system and instead rely on stochastic optimization, adaptive control logic, and machine learning methods \cite{brunton_2015_closed_loop_control, brunton2020machine}. Model-free control using machine learning (ML) was pioneered by Lee et al. \cite{lee1997neural} using Artificial Neural Networks (ANN), where the sensor-based
control law was learned from a known optimal full-information controller, with little loss in overall performance, achieving 20\% skin-friction drag reduction at \(Re=100\).

Reinforcement learning (RL) represents the latest advancement in ML control \cite{sutton2018reinforcement}. RL is a machine learning paradigm that is often considered the closest to human learning, as it approaches learning through trial and error, with successful actions being rewarded and errors penalized \cite{sutton2018reinforcement}. The RL agent is trained by observing the environmental state \(s_t\), responding with an appropriate action \(a_t\), and receiving a reward \(r_{t+1}\), with the objective of maximizing the expected global reward \(\mathcal{R}\) accumulated over a defined number of interactions (steps). The mapping from observed states to the agent's respective actions is referred to as a policy. In the stochastic case, the policy maps a state \(s_t\) to a probability distribution over the action space and is  denoted by \(\pi(\,\cdot \mid s_t)\).  

Early applications of RL to active separation control include the work of Guéniat et al. \cite{gueniat2016statistical}, who applied an RL-based control strategy to reduce drag around a bluff body in a two-dimensional (2D) laminar flow field, reporting a recirculation region reduction of approximately 15\%. The concept was later picked up by Rabault et al. \cite{rabault2018drl_afc, rebault_second}, whose work provided further proof of concept for future applications. In Ref. \cite{rabault2018drl_afc}, DRL coupled with ASC was applied as a method to develop a control strategy for two synthetic jets on a cylinder by varying the injected mass flow rate at \(Re=100\), with a reported drag reduction of 8\%. An extension by Tang et al. \cite{tang2020robust} investigated a similar set-up, with two additional synthetic jet actuators, at varying Reynolds numbers of up to \(Re=400\), reporting a drag reduction of up to 38.7\%.
Another follow up study by Ren et al. \cite{ren2021drl_turbulent} investigated the same set-up at a higher Reynolds number of \(Re=1000\) and confirmed the effectiveness of the implementation of the DRL-controller, with a reported drag reduction of 30\%, noting a longer training time needed for convergence. Further implementations include the work of Fan et al. \cite{fan2020reinforcement}, who applied this approach to discover control strategies for turbulent cylinder flow
with two fast-rotating smaller control cylinders in both experimental and simulation set-ups; followed by Shimomura et al. \cite{shimomura2020distributeddrl} and Zheng et al. \cite{zheng2024transformer}, both of whom implemented DRL for flow separation control over airfoils. Most recent publications include works by Font et al. \cite{Font_2025}, who focused on turbulent separation bubble (TSB) reduction, reporting a reduction of 9.0\% with DRL versus 6.8\% with periodic forcing; as well as Fang et al. \cite{fang_2026}, who implemented DRL-based closed-loop control to reduce friction drag in a turbulent boundary layer at a Reynolds number of \(Re_{\tau}=1196\), achieving a relative drag reduction of 6.7\%, almost three times that of the best open-loop case (2.3 \%).   

Despite recent progress, most RL-based separation-control studies have been conducted in numerical environments or at comparatively low Reynolds numbers. Consequently, the feasibility of training and deploying an RL controller directly within a fully turbulent experiment remains insufficiently established. Here, PPO is integrated into a wind-tunnel experiment and trained online to control separation over a one-sided diffuser. The controller uses the signal from a single bidirectional wall-shear-stress sensor as its state input and commands three in-phase pulse-jet actuators through binary valve actions. This implementation is intended to validate that a model-free control policy can be learned directly from interactions with a physical turbulent flow using sparse sensing and low-dimensional actuation. To investigate how temporal credit assignment affects the efficiency of online learning, two return formulations are compared: a conventional discounted return accumulated over the full trajectory and a finite-horizon return defined on the order of the characteristic flow-response time. The selected closed-loop controller is subsequently benchmarked against optimized periodic open-loop actuation.


\section{Methods}
\subsection{Experimental Setup}
\label{experimentalsetup}
The experiments were conducted in a closed-loop subsonic wind tunnel at a nominal free-stream velocity of \(U_{\infty} = 20\,\mathrm{m\,s^{-1}}\) at \(20\,^{\circ}\mathrm{C}\), equipped with a one-sided diffuser with a deflection angle of \(\alpha = 20^{\circ}\) and a ramp length of \(L = 337\,\mathrm{mm}\). The fully turbulent flow is characterized by a Reynolds number of \(Re_{\theta} = 1000\), based on a momentum thickness of \(\theta \approx 0.8\,\mathrm{mm}\).

\begin{figure}[tbp]
    \centering
    \begin{subfigure}[b]{0.4\linewidth}
        \centering
        \includegraphics[width=\linewidth]{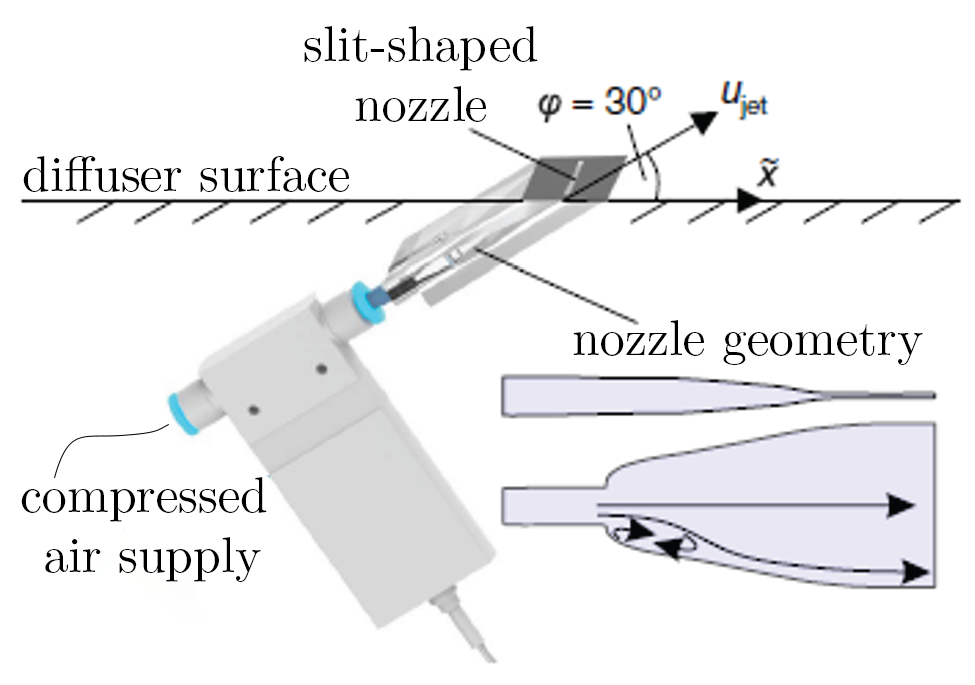}
        \caption{}
        \label{fig:PJA_geom}
    \end{subfigure}
    \begin{subfigure}[b]{0.4\linewidth}
        \centering
        \includegraphics[width=\linewidth]{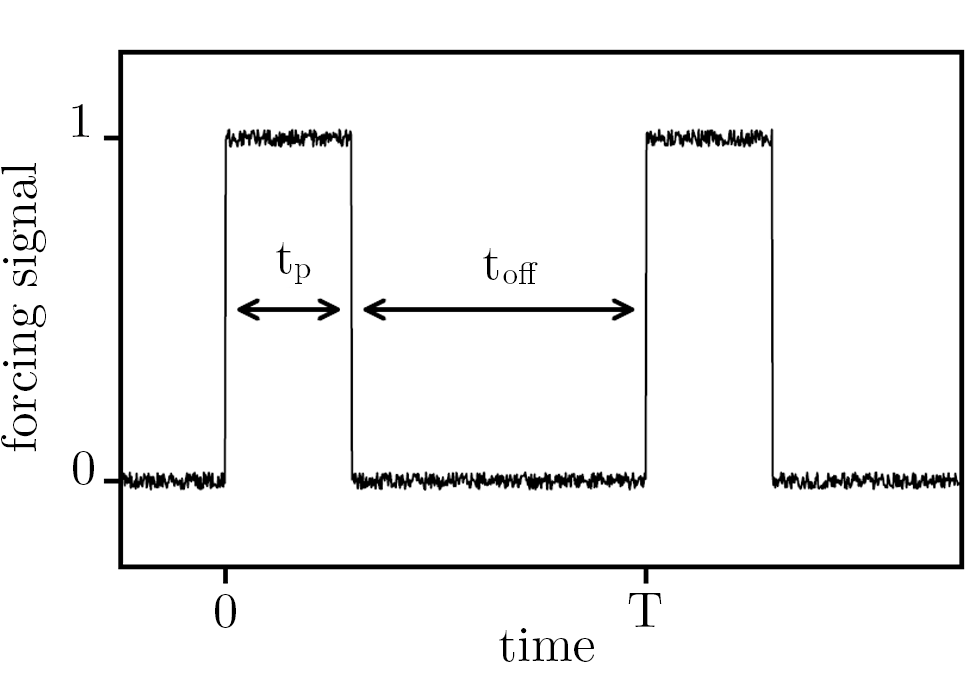}
        \caption{}
        \label{fig:PJA_sig}
    \end{subfigure}
    \caption{PJAs used for ASC: positioning in relation to the diffuser surface (a) \cite{steinfurth_weiss_2022} and a sequence of an exemplary forcing signal (b) \cite{loeffler2023}.}
    \label{fig:PJA}
\end{figure}

The geometry of the diffuser leads to pressure-induced flow separation \cite{loeffler2023, steinfurth_weiss_2022}. In this set-up, mass-flow injection is handled through three pulse jet actuators (PJAs) placed side-by-side near the top of the diffuser ramp upstream of the natural separation line, at $x\approx \SI{65}{mm}$ with a spacing of \SI{2}{mm} and an emission angle of \SI{30}{\degree}, representing an effective operating condition (Fig. \ref{fig:PJA_geom}). Connected to an external compressed air supply, these actuators operate by intermittently injecting fixed amounts of fluid through slit-shaped outlets with cross-sectional area of $38 \times 0.5\,\si{\milli\metre}$ into a cross-flowing boundary layer, where they increase the momentum flux \cite{steinfurth_2021}. All three actuators were operated in-phase, meaning the same control signal to open/close the valve was applied to all three simultaneously (Fig.\ref{fig:PJA_sig}).
In this set of experiments, the mass flow rate passed by a mass flow controller directly to the PJAs was preset to $\sim 5.02\cdot 10^{-4}\,\mathrm{kg/s}$, corresponding to a mean jet velocity of $u_{jet} = \SI{3.7}{\metre\per\second}$ ($\sim 0.185\, U_\infty$).  With this, the momentum coefficient, defined as the ratio of added momentum to free-stream momentum \cite{greenblatt_wygnanski_2000}, is obtained as $c_\mu \approx 0.003$. 

A detailed analysis of the flow structures generated with the PJAs is provided in a series of articles by Steinfurth \& Weiss \cite{steinfurth_2021, Steinfurth_Weiss_2020, Steinfurth_Weiss_2021, steinfurth_weiss_2022}, with \cite{steinfurth_weiss_2022} in particular focusing on the optimization of the forcing signal and, accordingly, its characteristic duty cycle (DC). 

More specifically, \cite{steinfurth_weiss_2022} examines the pulse duration \(t_p\) and the interval between successive pulses \(t_{off}\). Together, these parameters define the actuation period $T=t_\mathrm{p}+t_\mathrm{off}$, while the duty cycle is given by $DC=t_\mathrm{p}/T$ (see Fig. \ref{fig:PJA_sig}). The experiments were conducted using a closely related configuration. Although $DC=0.5$ can be considered the standard forcing signal in the literature, this work indicated that, for a given mass flow rate, the control authority can be increased by lowering the duty cycle \cite{steinfurth_weiss_2022}.

\begin{figure}[tbp]
    \centering
    \begin{subfigure}[c]{0.48\linewidth}
        \centering
        \includegraphics[width=\linewidth]{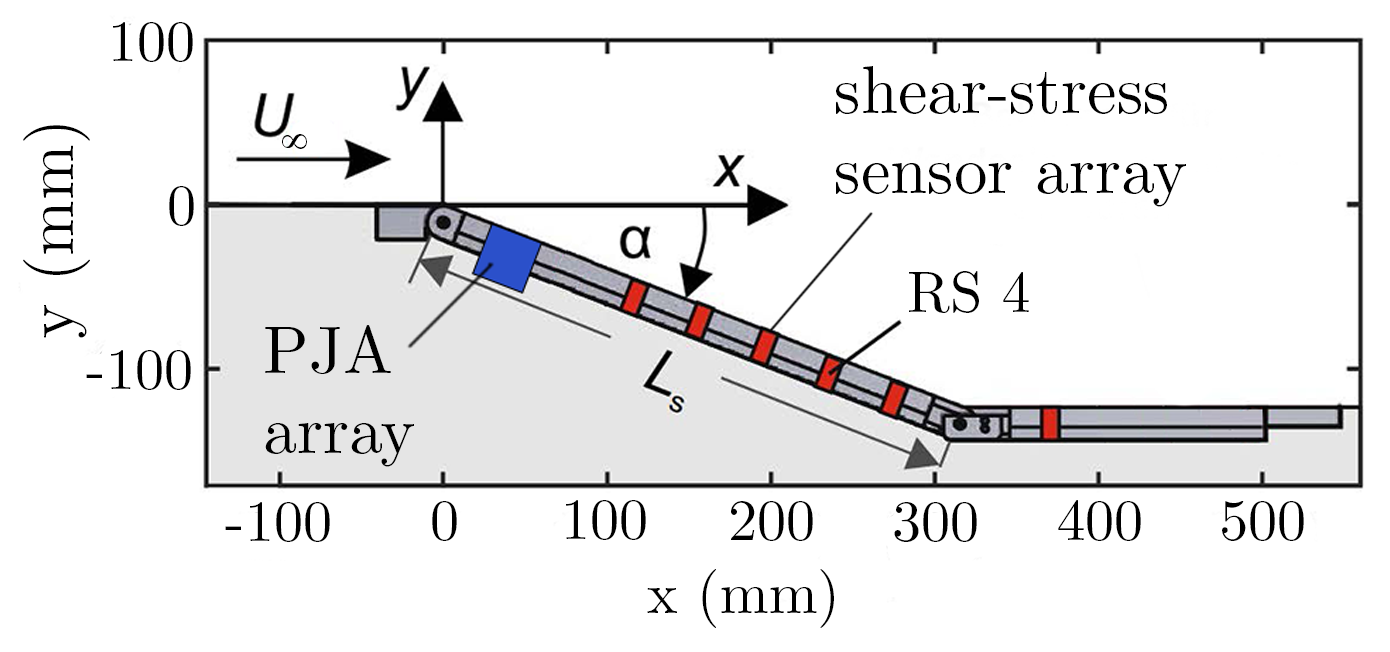}
        \caption{}
        \label{fig:setup_geometry}
    \end{subfigure}
    \begin{subfigure}[c]{0.48\linewidth}
        \centering
        \includegraphics[width=\linewidth]{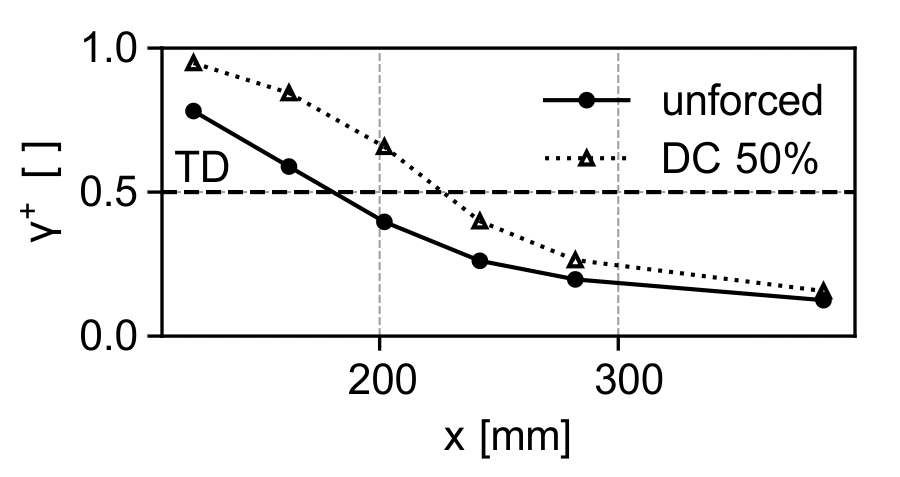}
        \caption{}
        \label{fig:setup_gamma}
    \end{subfigure}
    \caption{Actuator and sensor set-up in the half-diffuser test section (a), adapted from \cite{weiss2017mems}, and forward-flow fraction \(\gamma^+\) along the diffuser ramp in the unforced state and under periodic excitation (b).}
    \label{fig:setup}
\end{figure}

Five bi-directional MEMS colorimetric shear-stress sensors were placed flush along the ramp, with one additional sensor installed at the foot of the ramp, to assess the effectiveness of actuation in reattaching the flow to the wall surface (Fig. \ref{fig:setup_geometry}). These sensors are sensitive to the flow direction, which allows one to track the reverse flow at the surface of the diffuser wall. Here, a micro beam suspended over a micrometer-scale cavity is heated by an electric current, and the thermal wake is measured with two additional lateral detector beams acting as resistance thermometers \cite{weiss2017mems}. 

At each sensor location, the dimensionless local forward flow fraction \(\gamma^+\) is determined as the percentage of positive recorded values relative to the total number of recorded values. Therefore, a local forward flow fraction \(\gamma^+=1\) indicates fully attached flow at the measurement location, while \(\gamma^+ \geq 0.5\) characterizes predominantly attached flow. Figure \ref{fig:setup_gamma} shows the distribution of the local forward flow fraction of the flow field along the sensor array in the natural state (unforced) and under periodic actuation (DC=50\%, $t_p=t_{off}=\SI{30}{\milli\second}$). The adverse pressure gradient leads to boundary layer separation, marked by $\gamma^+=0.5$ \cite{loeffler2023, steinfurth_weiss_2022}. The transitory detachment point (TD) of the natural flow field, past which the recorded shear stress becomes negative on average, can be located at $x\approx 180\,\mathrm{mm}$, while under actuation, it drifts further downstream to $x\approx 230\,\mathrm{mm}$. For the DRL implementation, reference sensor 4 (RS4) has been chosen as the environmental state evaluator since it is placed closest to the transitory detachment point under actuation.

\subsection{DRL Control Framework}
\label{DRL}

The approach used in this study to optimize the policy follows the proximal-policy optimization (PPO) approach due to its sample efficiency and robustness. PPO is an on-policy algorithm, meaning the learning machine explores the environment by executing actions according to the latest version of its policy, which is then updated using stochastic gradient descent within a restricted trust region limited by the clip factor $\epsilon$ \cite{Schulman2017ProximalPO}. The algorithm is implemented in Python using the machine learning package PyTorch \cite{github}.

PPO is an actor-critic algorithm, where the actor explicitly represents the stochastic policy \(\pi_{\theta}(\,\cdot \mid s_t)\) and outputs the probability of selecting action \(a_t\) in state \(s_t\), and the critic represents the state-value function \(\mathcal{V}\) used to update the actor's weights \(\xi\) \cite{Konda1999actorcritic}. 

In the present experimental set-up, the state \(s_t\) is defined as the voltage measured at RS4; note that a positive voltage indicates forward-directed flow. The sensor voltage output can be mapped to wall shear stresses through a calibration polynomial. However, the additional computation would increase the real time delay and is therefore spared. For each state \(s_t\), the actor outputs a scalar probability \(p_t\in(0,1)\), obtained by applying a sigmoid function to the scalar network output. This probability parametrizes a Bernoulli distribution over the binary action space \(a_t\in\{0,1\}\), such that

\begin{equation}
P(a_t=1\mid s_t)=p_t,
\qquad
P(a_t=0\mid s_t)=1-p_t.
\end{equation}

An action is sampled from this distribution at each interaction. It is important to note that values of \(p_t\) express the probability of opening the valves rather than intermediate valve positions. The sampled action \(a_t=1\) commands all three magnetic valves to open simultaneously, whereas \(a_t=0\) commands them to remain closed. Finally, the instantaneous reward is determined by the sign of the state \(s_{t+1}\), measured after action \(a_t\) is executed:

\begin{equation}
r_{t+1} = \operatorname{sgn}\!\left(s_{t+1}\right)
=
\begin{cases}
+1, & s_{t+1} \geq 0,\\
-1, & s_{t+1} < 0.
\end{cases}
\end{equation}

A positive sensor signal, corresponding to forward-directed wall shear-stress, is therefore rewarded, whereas a negative signal, indicating reverse flow, is penalized.

Each interaction with the environment produces a quadruple of the form \((s_t,a_t,s_{t+1},r_{t+1})\). Each learning episode consists of 1000 interactions, with the collected batch of data leveraged in the agent update loop using the Adam optimizer. This requires returns \(\mathcal{R}\) and advantages \(\mathcal{A}\). 

The policy update is based on the estimated advantage, which compares the return assigned to action \(a_t\) with the return expected by the critic from the current state \(s_t\):

\begin{equation}
\label{eq:advantage}
\widehat{A}_t
=
\widehat{R}_t
-
V_{\psi}(s_t),
\end{equation}

where \(\widehat{R}_t\) denotes the return assigned to action \(a_t\), and \(V_{\psi}(s_t)\) is the corresponding value predicted by the critic with parameters \(\psi\). A positive advantage increases the probability of the sampled action, whereas a negative advantage decreases it, subject to the clipping constraint of PPO. Consequently, both the instantaneous reward definition and the temporal interval over which rewards are combined determine the credit assigned to an action \cite{guastoni_2023,fang_2026,mueller2026optimizing}.




Two return definitions are investigated. The first agent, hereafter referred to as the infinite-horizon agent, uses a discounted return accumulated over all remaining interactions in the trajectory:

\begin{equation}
\label{eq:return_infinite}
\widehat{R}^{\mathrm{IH}}_t
=
\sum_{k=1}^{T-t}
\gamma^{\,k-1}
r_{t+k},
\end{equation}

where \(\gamma=0.995\) is the discount factor. Although each trajectory is finite, the term infinite-horizon agent is used to distinguish this full-trajectory return from the explicitly truncated formulation introduced below.

The second agent, hereafter referred to as the finite-horizon agent, uses only the first \(N_\mathrm{c}\) instantaneous rewards following an action:

\begin{equation}
\label{eq:return_finite}
\widehat{R}^{\mathrm{FH}}_t
=
\sum_{k=1}^{N_t}
r_{t+k},
\qquad
N_t
=
\min\!\left(N_\mathrm{c},T-t\right),
\end{equation}

where \(N_\mathrm{c}=4\) and no discounting is applied \((\gamma=1)\). With an interaction time of \(\Delta t=6\,\mathrm{ms}\), the nominal four-step horizon corresponds to \(N_\mathrm{c}\Delta t=24\,\mathrm{ms}\), which is of the same order as the characteristic convective and transient wall shear-stress response times previously identified for this diffuser configuration \cite{steinfurth_weiss_2022}. The truncated return therefore substantially narrows temporal credit assignment relative to the full-trajectory formulation. 

The hyperparameters used are listed in Table \ref{tab:hyperparameters}.

A clip factor of \(\epsilon = 0.1\) has been shown to offer more stable long-term learning performance, as it is more restrictive than the standard value of 0.2 and therefore counters wide short-term oscillations in policy updates more effectively. The entropy bonus was factored out due to the inherently high stochasticity of the environment, which already ensures sufficient exploration of the state space. To ensure neutral starting conditions in each episode, during the update phase of both models, the flow actuation is maintained according to the same policy as that applied during trajectory collection, using a deep copy of the actor network.  

\begin{table}[hbt!]
\centering
\caption{\label{tab:hyperparameters}
List of hyperparameters used for the two investigated training approaches.}
\begin{tabular}{p{0.3\textwidth} p{0.32\textwidth} p{0.32\textwidth}}
\hline
\textbf{Parameter}
& \textbf{Infinite-horizon agent}
& \textbf{Finite-horizon agent} \\
\hline
Discount factor \(\gamma\) & 0.995 & 1 \\
Step duration \(dt\) & 6 ms & 6 ms \\
Trajectory length \(T\) & 1000 & 1000 \\
Return upper limit & \(T\) & 4 steps \\
Epsilon clip \(\epsilon\) & 0.1 & 0.1 \\
Update epochs & 100 & 100 \\
Training episodes & 100--500 & 100--500 \\
Actor learning rate & \(1 \cdot 10^{-3}\) & \(1 \cdot 10^{-3}\) \\
Critic learning rate & \(1 \cdot 10^{-4}\) & \(1 \cdot 10^{-4}\) \\
Actor/Critic inputs & 1 & 1 \\
Actor/Critic outputs & 1 & 1 \\
Number of hidden layers & 2 & 2 \\
Number of hidden neurons & 64 & 64 \\
\hline
\end{tabular}
\end{table}
\section{Results}
This section first examines the online learning of separation-control policies in the physical wind-tunnel experiment (\ref{sec:results_models}) and subsequently benchmarks the selected controller against periodic open-loop AFC (\ref{comparison}). Two agents are compared to assess how temporal credit assignment affects the efficiency and stability of policy identification in the turbulent flow. The infinite-horizon agent uses a discounted return accumulated over the remaining trajectory, whereas the finite-horizon agent uses a truncated return defined on the order of the characteristic flow-response time. The latter formulation reduces the influence of delayed rewards that are less directly attributable to an individual action. The comparison therefore serves both to demonstrate that useful control strategies can be learned directly from online interactions with the experiment and to determine which return formulation is more suitable for subsequent deployment.

\subsection{Closed-loop control using DRL}
\label{sec:results_models}
\begin{figure}[tbp]
    \centering
    \makebox[\textwidth][c]{%
        \begin{subfigure}[b]{0.48\linewidth}
            \centering
            \includegraphics[width=\linewidth]{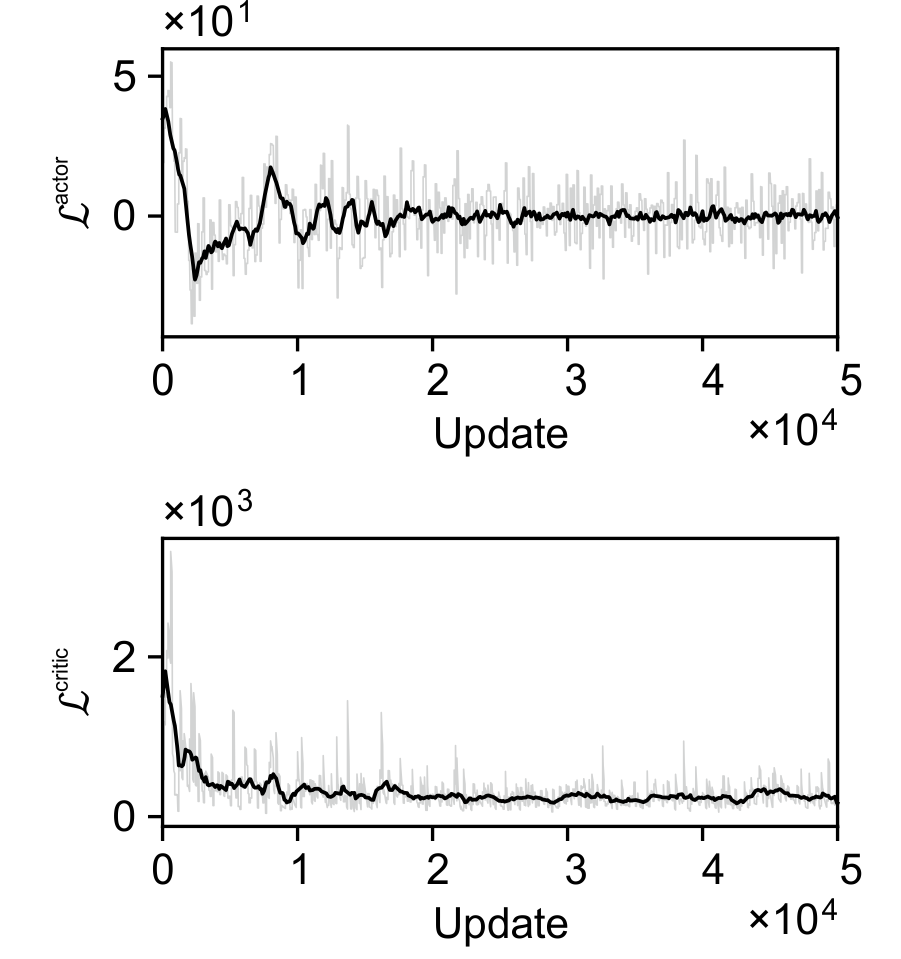}
            \caption{}
            \label{fig:losses_infinite}
        \end{subfigure}
        \hfill
        \begin{subfigure}[b]{0.48\linewidth}
            \centering
            \includegraphics[width=\linewidth]{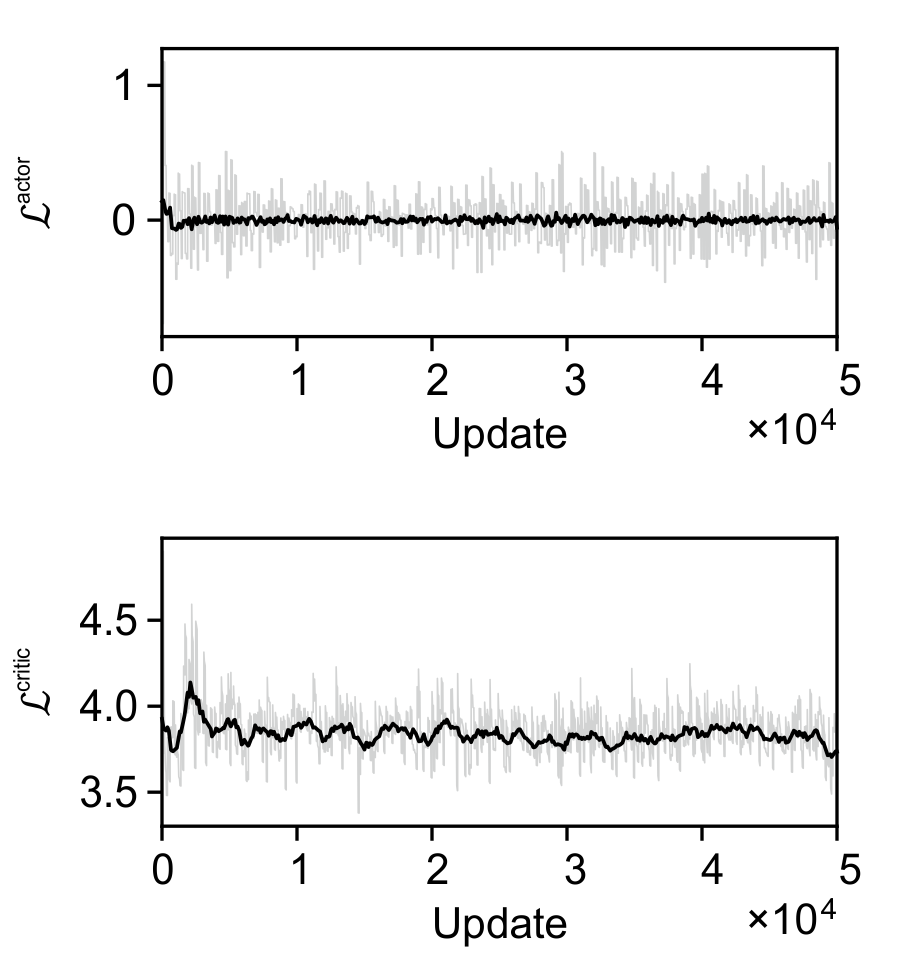}
            \caption{}
            \label{fig:losses_finite}
        \end{subfigure}%
    }
    \caption{Evolution of actor and critic losses over updates for the infinite-horizon agent (a) the finite-horizon agent (b); black lines represent moving averages over a window of 1000 updates while gray lines indicate loss per update}
    \label{fig:losses_model1_model2}
\end{figure}

Figure \ref{fig:losses_model1_model2} illustrates the evolution of actor and critic losses for the two selected agents over 500 episodes, each comprising 100 update iterations. It is to be expected that in early episodes, both agents experience high losses for both the critic and the actor networks, as the mappings of inputs to outputs are not yet shaped. Note that for different return definitions, both $\mathcal{L}^{actor}$ and $\mathcal{L}^{critic}$ have significantly different value ranges; this scale effect is, however, distinct from the credit-assignment effect. Moreover, due to the simplified Monte Carlo advantage definition, which is sensitive to the high variance present in a turbulent flow, losses cannot be expected to fully converge towards zero for either agent; therefore, slight oscillations around low values are to be tolerated.

Both agents start their training at relatively high losses (Fig. \ref{fig:losses_model1_model2}). The finite-horizon agent reaches a stable low-duty-cycle operating regime within approximately 20 episodes, whereas the infinite-horizon agent continues to exhibit substantial duty-cycle variation throughout the investigated 500-episode training run. Faster convergence of the agent based on the truncated return definition may be explained by a significantly shorter local reward observation window and, therefore, a tighter credit assignment interval. Consequently, the resulting advantage estimates are expected to exhibit lower variance, which leads to a more stable and effective policy update.

\begin{figure}[htbp]
    \centering
    \makebox[\textwidth][c]{%
        \begin{subfigure}[b]{0.5\textwidth}
            \centering
            \includegraphics[width=\linewidth]{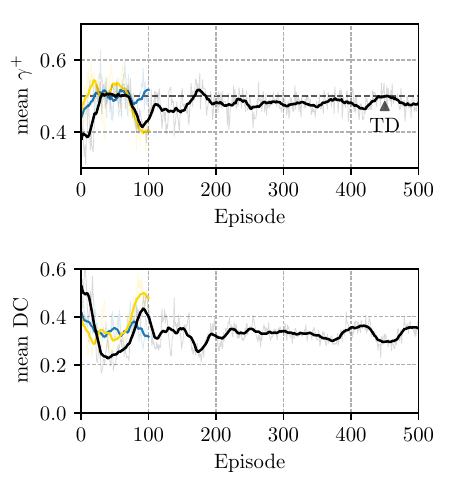}
            \caption{}
            \label{fig:no_horizon_dc_gamma}
        \end{subfigure}
        \hfill
        \begin{subfigure}[b]{0.5\textwidth}
            \centering
            \includegraphics[width=\linewidth]{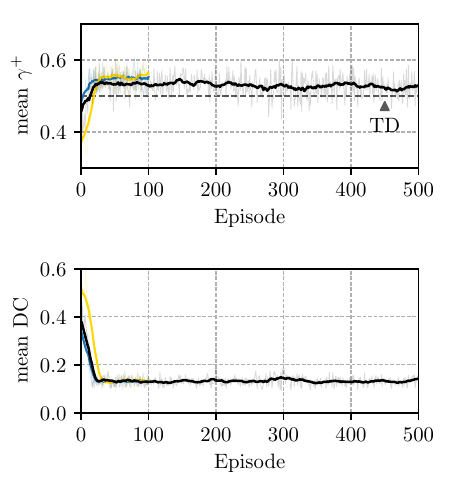}
            \caption{}
            \label{fig:short_horizon_dc_gamma}
        \end{subfigure}%
    }
    \caption{Evolution of the training metrics for infinite-horizon agent (a) and finite-horizon agent (b) over 500 training episodes, note that for readability, the black and colored curves are smoothed using a running average over a 20 episode window; blue and yellow curves represent repetition runs (terminated after 100 episodes).}
    \label{fig:learning_curves_model1_model2}
\end{figure}

Figure \ref{fig:learning_curves_model1_model2} illustrates the respective learning curves of the infinite-horizon agent (\ref{fig:no_horizon_dc_gamma}) and the finite-horizon agent (\ref{fig:short_horizon_dc_gamma}). The performance of the developed control strategies is assessed based on the mean local forward flow fraction \(\gamma^+\) at the reference sensor RS4, while the corresponding duty cycle is displayed in the bottom panels. Since DRL control is non-periodic, the mean duty cycle is defined as the ratio of the number of steps during which the valves were open to the total number of steps in the trajectory; similarly, the mean \(\gamma^+\) is defined over the full trajectory length in each training episode. To assess the repeatability of training, two further runs (yellow and blue curves) were executed under the same conditions, but terminated after 100 episodes.

For both agents, the initial policy is approximately equivalent to an unbiased probability distribution over a binary action space: at any input state, the probability of an open command to the valves is $p\approx0.5$. As a result, both agents start with a mean duty cycle of around 50\%, as evident from Fig. \ref{fig:learning_curves_model1_model2}. Since positive states are rewarded and negative states are penalized, the evolution of the mean \(\gamma^+\) also roughly translates to the evolution of the cumulative return $\mathcal{R}$, which influences the learning signals of both agents, albeit to a different extent depending on the exact definition. As established through the baseline measurements summarized in Fig. \ref{fig:setup_gamma}, with a DC of 50\%, \(\gamma^+\) in RS4 remains below the value of 0.5, which means that most recorded stresses are negative and the flow is, on average, detached. Initially, both agents successfully recognize these states as undesirable and update their policies in order to escape them. Consequently, both agents reduce the duty cycle during the first few training episodes. The agent trained with infinite-horizon returns follows a more exploratory path and temporarily increases the duty cycle, as a result, obtaining lower forward flow fractions per episode and subsequently lower returns (\ref{fig:no_horizon_dc_gamma}). In contrast, the agent trained with the finite-horizon return reaches its performance optimum around the 20th training episode and remains close to this operating point while further fitting its policy to the collected data (\ref{fig:short_horizon_dc_gamma}). Ultimately, the agent trained on the truncated return signal shows better performance and achieves a steady, favorable mean duty cycle consistent with previous open-loop measurements in a similar set-up \cite{steinfurth_weiss_2022}. 

Over the course of training, policies usually become progressively less explorative, as the update rule encourages them to exploit rewards that have already been identified \cite{openai_spinningup_ppo}. Accordingly, it is natural to expect the control policy to become progressively more refined and to converge towards a stable strategy. However, the infinite-horizon agent's learning behavior is characterized by persistent fluctuations that may be attributed to higher exploration. Although early exploration is common and can be beneficial, as it allows the agent to sample different state-action combinations, the exploratory behavior observed here persists throughout most of the training. Given binary action space and sign-based reward, the later duty-cycle variation is unlikely to reflect meaningful exploration. It can be more plausibly attributed to noisy advantage estimates and less direct credit assignment under the untruncated return. The infinite-horizon reward signal definition renders the consequences of individual actions less tangible to the agent, limiting its potentially achievable performance. As a result, the mismatch between the approximation of performance and the actual collected global reward becomes significant, and large advantages \(\mathcal{A}\) are obtained, leading to large steps of policy update. 

The degree of policy certainty can be tracked through the entropy $\mathcal{H}$, defined as follows:
\begin{equation}
\label{entropydef}
\mathcal{H}(\pi(\cdot \mid s_t)) = - \sum_{a \in \mathcal{A}} \pi(a \mid s_t)\,\log \pi(a \mid s_t).
\end{equation}

With a binary action space, (\ref{entropydef}) can be simplified to the following form:
\begin{equation}
\label{entropysimple}
\mathcal{H}_{step}(p(s_t)) = - \left[ p \log p + (1 - p)\log(1 - p) \right].
\end{equation}

\begin{figure}[htbp]
    \centering
    \makebox[\textwidth][c]{%
        \begin{subfigure}[b]{0.5\textwidth}
            \centering
            \includegraphics[width=\linewidth]{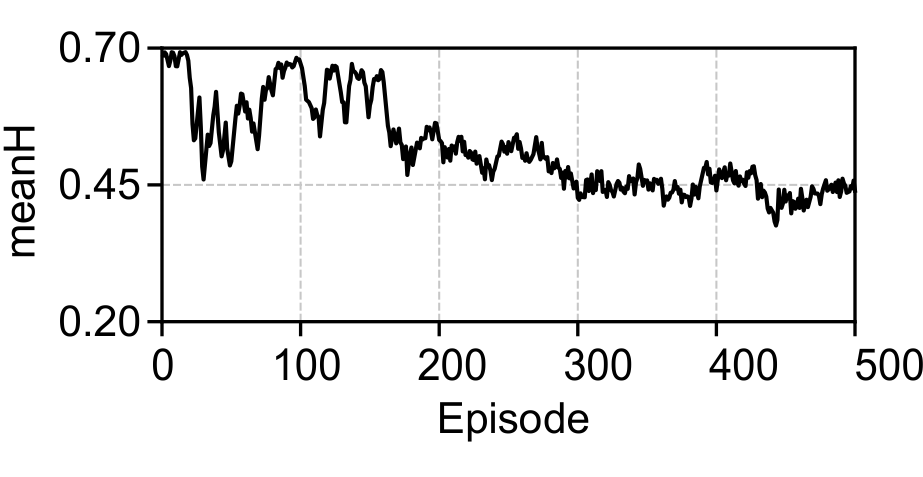}
            \caption{}
            \label{fig:no_horizon_entropy}
        \end{subfigure}
        \begin{subfigure}[b]{0.5\textwidth}
            \centering
            \includegraphics[width=\linewidth]{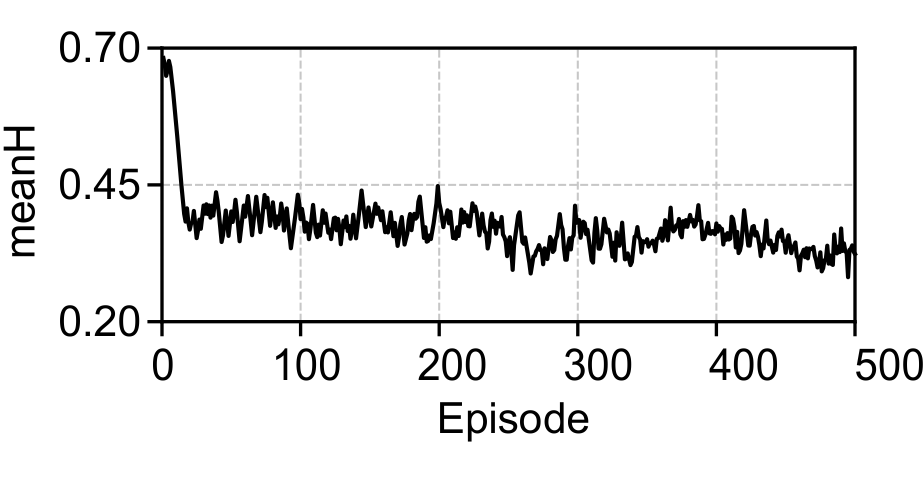}
            \caption{}
            \label{fig:short_horizon_entropy}
        \end{subfigure}%
    }
    \caption{Entropy evolution over the training episodes for the agent based on conventional infinite-horizon returns (a) and for the agent based on returns adjusted to convective time (b).}
    \label{fig:entropies}
\end{figure}

As evident from the evolution of the mean entropy over the training episodes shown in Figure \ref{fig:entropies}, both agents become increasingly deterministic throughout the course of training. However, compared to agent trained with truncated rewards, the infinite-horizon agent's policy remains less certain of its actions with comparably still high entropy. The mean episode entropy of the finite-horizon agent rapidly drops to a value of $\mathcal{H}=0.4$ in the first 20 training episodes and steadily decreases further to a value of $\mathcal{H}=0.3$. Combined with the low valve-open probabilities and mean duty cycle, the reduced entropy indicates an increasingly deterministic preference for the closed-valve action.

\begin{figure}[htbp]
    \centering
    \begin{subfigure}[b]{0.49\textwidth}
        \centering
        \includegraphics[width=\linewidth]{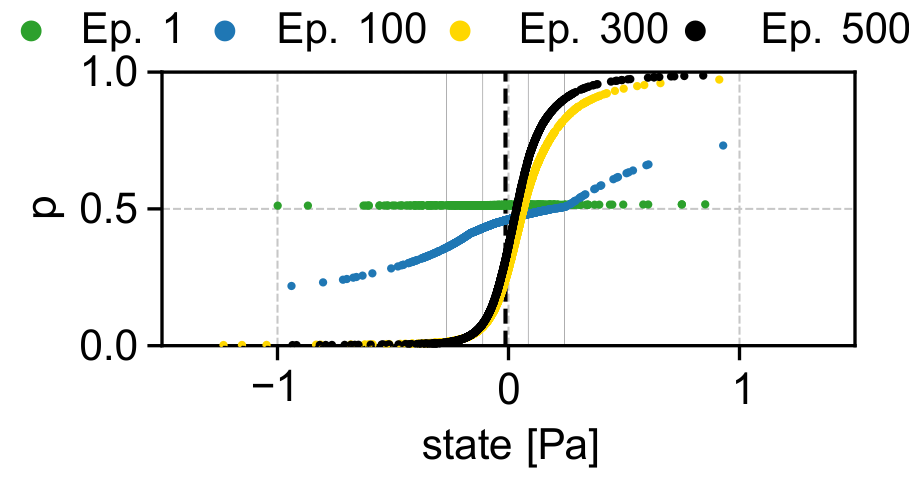}
        \caption{}
        \label{fig:prob_model1}
    \end{subfigure}
    \hfill
    \begin{subfigure}[b]{0.49\textwidth}
        \centering
        \includegraphics[width=\linewidth]{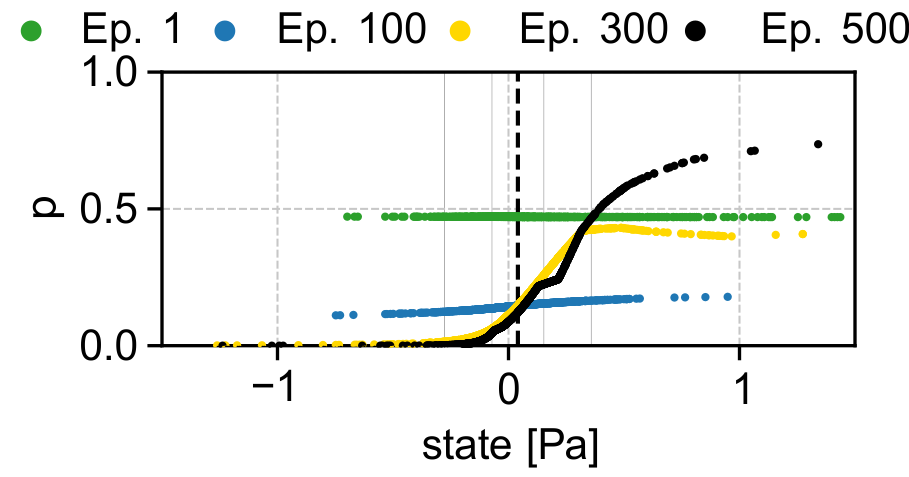}
        \caption{}
        \label{fig:prob_model2}
    \end{subfigure}
    \caption{Policy plots over the course of training for the agent based on conventional infinite-horizon returns (a) and for the agent based on returns adjusted to convective time (b): probability of opening the valves as a function of the state, with the mean episode state indicated by the dashed vertical line, the 90\% range by the hatched area, and the 50\% range by the gray area at Episode 500, together with the policies at Episodes 300, 100 and 1. For ease of interpretation, the sensor output voltage was converted to wall shear stress.}
    \label{fig:prob_models}
\end{figure}

Figure \ref{fig:prob_models} shows the states that were visited during the last training episode, as well as the policies executed during episodes 100, 300, and 500 (final episode). Although both agents have 50\% of all visited states lying within the same narrow interval around zero, the infinite-horizon agent succeeds in keeping 90\% of its visited states more tightly concentrated around the same value. A tighter distribution of visited states indicates smaller fluctuations at RS4 and hence more stable $\gamma^+$ within a training episode, implying that the agent based on infinite-horizon return is learning with respect to long term performance, which is consistent with the expectation given its return definition. However, the computed probabilities span the full range from 0 to 1, consistent with the larger duty cycle observed for the infinite-horizon agent. By contrast, for 90\% of the states visited by the finite-horizon agent, the predicted valve-open probability remains below 0.5. In other words, it learns to output duty cycles only in the range from 0\% to 50\% for the majority of states. 

Overall, the experiments demonstrate that PPO can identify a low-duty-cycle separation-control strategy directly from online interactions with a turbulent wind-tunnel flow. Both agents reduce the initially high actuation rate and improve the forward-flow fraction at the feedback-sensor location, confirming that useful control policies can be learned from a single sensor and binary valve commands. The comparison further shows that the temporal formulation of the return strongly affects the efficiency of this online learning process. The finite-horizon agent reaches a favorable operating regime within the early training episodes and subsequently maintains a stable low-duty-cycle strategy, whereas the infinite-horizon agent exhibits persistent policy and duty-cycle variations. The finite-horizon agent is therefore selected for comparison with periodic open-loop actuation.

\subsection{Comparison to open-loop periodic control}
\label{comparison}
To establish a better frame of reference for evaluating the learning results of the best performing DRL-controller, an additional sequence of measurements was taken using optimized open-loop control. These were performed under the same boundary conditions as those used during DRL-training. The open-loop control sequence was executed with a Duty Cycle of 12.5\% to closely match that of the trained DRL-controller. For the latter, the agent trained with truncated rewards was deployed after 100 training episodes, since it had demonstrated sufficient action-state certainty reflected in its entropy and constructed policy. Note that for the given configuration, optimal open-loop control is indeed achieved at this duty cycle: that is, the trained DRL-agent has discovered a closed-loop strategy closely matching the expectations for the best performing open-loop actuation in just about 10 minutes of flow interaction.

\begin{figure}[htbp]
    \centering
    \makebox[\textwidth][c]{%
        \begin{subfigure}[b]{0.5\textwidth}
            \centering
            \includegraphics[width=\linewidth]{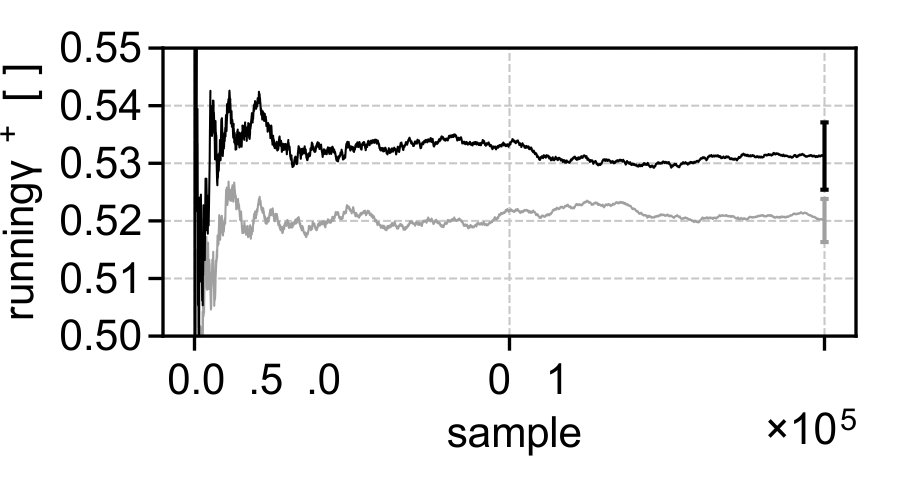}
            \caption{}
            \label{fig:comparison_final_a}
        \end{subfigure}
        \hfill
        \begin{subfigure}[b]{0.5\textwidth}
            \centering
            \includegraphics[width=\linewidth]{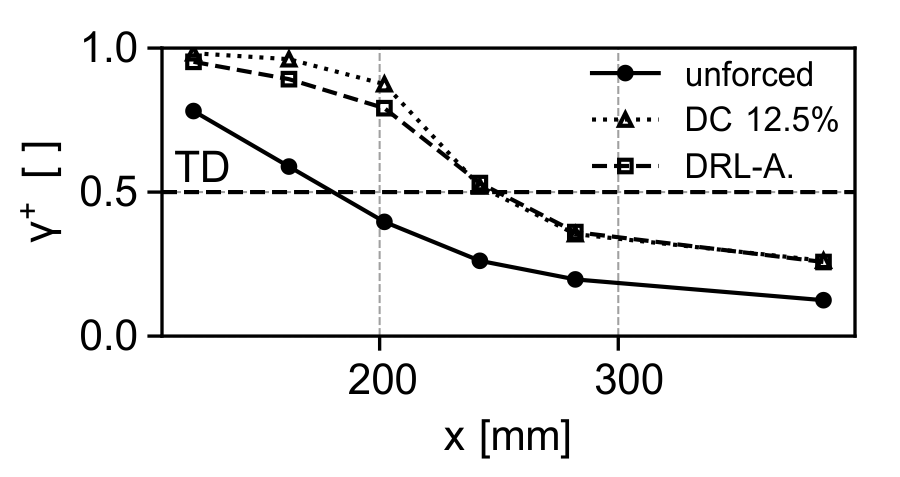}
            \caption{}
            \label{fig:ramp_inference}
        \end{subfigure}%
    }
    \caption{Convergence of the local forward-flow fraction \(\gamma^+\) measured at RS4 for the deployed finite-horizon agent (black) and periodic actuation at a duty cycle of 12.5\% (gray); error bars indicate the autocorrelation-corrected 95\% confidence intervals for both time series (a). Distribution of the local forward-flow fraction across the sensor array in the natural state (\(\bullet\)), under periodic actuation at a duty cycle of 12.5\% (\(\triangle\)), and under targeted actuation (\(\square\)) (b).}
    \label{fig:comparison_final}
\end{figure}

Figure \ref{fig:comparison_final} shows the cumulative running mean forward-flow fraction along a trajectory of $10^5$ time steps for both the open-loop controller and the DRL controller. The DRL control is shown to slightly outperform the open-loop control by a margin of approximately 1\% (Fig. \ref{fig:comparison_final_a}). Figure \ref{fig:ramp_inference} shows the forward flow fraction distribution measured throughout the sensor array. While both controllers perform well in the area of the separation bubble (indicated by the cutoff value of \(\gamma^+=0.5\)), periodic actuation yields a higher forward flow fraction closer to the PJA outlets in the region \(x=[0,200]\). This may be attributed to the environmental state being evaluated at a single sensor, RS4, rather than the full sensor array. The benefit of DRL-control is therefore limited to RS4, as further downstream at RS5 and RS6, both performance curves essentially merge. 

As argued in Ref. \cite{steinfurth_weiss_2022}, keeping \(t_{off}\) below a flow-specific separation time is essential to maintaining high control authority over the flow field. For stochastic DRL control, however, \(t_{\mathrm{off}}\) accumulated over consecutive valve closures exceeds the characteristic separation time. It is reasonable to believe that this has a negative effect predominantly in the upstream part of the diffuser where the injected disturbances are well separated. Further downstream, the pulsed jets have diffused to a greater extent, and the control authority is primarily governed by the mean momentum supply.

\section{Conclusions}

This work demonstrates the online training and deployment of a model-free separation controller in a fully turbulent wind-tunnel experiment. Proximal Policy Optimization was coupled directly to the experimental facility, with the policy receiving the signal from a single bidirectional wall-shear-stress sensor located near the transitory detachment point and commanding three in-phase pulse-jet actuators through binary valve actions. The control law was therefore learned from direct interactions with the wind-tunnel flow. The results show that reinforcement learning can serve as an experimental tool for identifying effective and interpretable actuation strategies under sparse sensing and low-dimensional actuation.

The comparison between the two return formulations identifies temporal credit assignment as an important design consideration for online experimental learning. Both agents reduced the initially high actuation rate and improved the forward-flow fraction at the feedback-sensor location. However, the finite-horizon agent reached a stable low-duty-cycle operating regime within approximately 20 episodes, corresponding to about two minutes of recorded flow interaction, whereas the agent using a discounted return over the full trajectory continued to exhibit substantial policy and duty-cycle variations. The results therefore indicate that restricting the return horizon to the order of the characteristic flow-response time can improve the efficiency and stability of policy identification in a noisy turbulent environment. Although direct comparisons of training times across different configurations and implementations must be interpreted cautiously, the short interaction time required here is consistent with the potential of training reinforcement-learning controllers directly in physical experiments \cite{Font_2025,fang_2026}.

When deployed after training, the finite-horizon controller achieved a forward-flow fraction at the feedback-sensor location approximately one percentage point higher than that obtained with optimized periodic forcing at a comparable mean duty cycle. This local improvement was accompanied by lower forward-flow fractions upstream, while the two controllers performed similarly farther downstream. The comparison highlights both the capability and the present limitation of the learned strategy: the controller discovered a state-dependent, non-periodic actuation pattern that cannot be represented by a single prescribed forcing period, but its objective was determined exclusively by the response at one sensor location. The resulting policy should therefore be interpreted as a locally optimized separation-control strategy rather than as a uniform improvement across the entire diffuser.

The present results suggest several directions for extending the approach. Incorporating signals from multiple wall-shear-stress sensors into the state representation could enable the controller to account for the spatial evolution of the separated flow and optimize performance across a larger portion of the diffuser. Independent control of the three actuators would further enlarge the action space and permit the discovery of spanwise-varying forcing strategies, as demonstrated in Ref.~\cite{Font_2025}. In addition, excessively long sequences of valve closures could be discouraged through a reward contribution associated with prolonged intervals approaching the characteristic separation time \(t_s\) or with increased wall-shear-stress fluctuations. Together, these developments could preserve the rapid online learning demonstrated here while extending reinforcement learning from localized separation mitigation toward spatially coordinated and increasingly autonomous active flow control.

\section*{Acknowledgments}
OpenAI ChatGPT was used to assist with language editing and to improve the readability and clarity of selected passages of the manuscript. All scientific content, analyses, interpretations, and conclusions were developed and verified by the authors, who take full responsibility for the manuscript.

\bibliography{refs}

\end{document}